\documentclass{article}

\usepackage{PRIMEarxiv}

\usepackage[utf8]{inputenc} 
\usepackage[T1]{fontenc}    
\PassOptionsToPackage{hyphens}{url}
\usepackage{hyperref}       
\usepackage{booktabs}       
\usepackage{amsfonts}       
\usepackage{nicefrac}       
\usepackage{microtype}      
\usepackage{lipsum}
\usepackage{fancyhdr}       
\usepackage{graphicx}       
\graphicspath{{media/}}     

\usepackage{tabularx}
\usepackage{booktabs}
\usepackage{ragged2e}

\usepackage{enumitem}

\usepackage{verbatim}       
\usepackage{xcolor}         
\usepackage{amsmath}
\usepackage{hyperref}
\usepackage{float}

\usepackage{multirow}

\usepackage{soul}

\usepackage{rotating}       

\usepackage{forest}
\usepackage{xcolor}
\usepackage{tikz}
\usetikzlibrary{arrows.meta, shapes, positioning}
\usepackage{geometry}
\usepackage{graphicx}
\usepackage{caption}
\usepackage{lscape}

\newcommand{\sunny}[1]{\textcolor{blue}{\textbf{Sunny:} #1}}

\title{The "4W+1H" of Software Supply Chain Security Checklist  for Critical Infrastructure}

\author{
  Liming Dong, Sung Une Lee, Zhenchang Xing, Muhammad Ejaz Ahmed\\
  Data61, CSIRO, Australia\\
     \AND
  Stefan Avgoustakis \\
  Google, Australia \\
}

\begin{document}
\maketitle

\begin{abstract}
The increasing frequency and sophistication of software supply chain attacks pose severe risks to critical infrastructure sectors, threatening national security, economic stability, and public safety. Despite growing awareness, existing security practices remain fragmented and insufficient, with most frameworks narrowly focused on isolated life cycle stages or lacking alignment with the specific needs of critical infrastructure (CI) sectors. 
In this paper, we conducted a multivocal literature review across international frameworks, Australian regulatory sources, and academic studies to identify and analyze security practices across the software supply chain, especially specific CI sector.
Our analysis found that few existing frameworks are explicitly tailored to CI domains. 
We systematically leveraged identified software supply chain security frameworks, using a "4W+1H" analytical approach, we synthesized ten core categories (what) of software supply chain security practices, mapped them across life-cycle phases (when), stakeholder roles (who), and implementation levels (how), and examined their coverage across existing frameworks (where).
Building on these insights, the paper culminates in structured, multi-layered checklist of 80 questions designed to relevant stakeholders evaluate and enhance their software supply chain security. 
Our findings reveal gaps between framework guidance and sector-specific needs, highlight the need for integrated, context-aware approaches to safeguard critical infrastructure from evolving software supply chain risks.
\end{abstract}




\section{Introduction} \label{sec:introduction}
Software now forms the backbone of modern critical infrastructure. From energy grids and healthcare systems to financial networks and transportation, nearly every essential service depends on complex layers of software, cloud platforms, and digital supply chains. However, as software dependencies grow, so do their vulnerabilities. Recent incidents such as SolarWinds and Codecov have shown how a single compromised component can ripple across thousands of systems, disrupting services and eroding public trust at a global scale. These events demonstrate that software security can no longer be viewed as an isolated technical task, it is a systemic challenge that spans the entire software life cycle\cite{Williams2025}, from code development and build environments to deployment and maintenance.

Traditional approaches to software security, such as static analysis, vulnerability scanning, and post-deployment patching, are proving inadequate against the evolving threat landscape\cite{WilliamsHZ25}. Attackers now exploit complex, multi-stage supply chains by compromising build systems, injecting malicious code into trusted components, or leveraging insider access to bypass controls. These tactics challenge\cite{melara2022what} conventional notions of software integrity and reveal a pressing need for end-to-end security assurances.

This concern is particularly emerging in critical infrastructure sectors\cite{paper1}, where software failures can have cascading impacts on national security, economic continuity, and public safety. From energy grids and water systems to healthcare platforms and financial services, CI operations increasingly depend on complex software ecosystems involving numerous suppliers, dependencies, and third-party components. In Australia, the 2023–2030 Cyber Security Strategy\footnote{2023-2030 Australian Cyber Security Strategy. \url{https://www.homeaffairs.gov.au/about-us/our-portfolios/cyber-security/strategy/2023-2030-australian-cyber-security-strategy}} explicitly identifies software supply chain security as a national priority, calling for coordinated and systematic efforts to secure digital infrastructure.

Multiple software supply chain security frameworks have emerged, such as NIST's Secure Software Development Framework (SSDF), SLSA, and OWASP SAMM—most focus narrowly on specific life cycle phases or general purpose use cases. Our review finds that few existing frameworks are tailored to the unique operational constraints, regulatory requirements, and risk profiles of CI sectors, particularly in the Australian context. This mismatch creates gaps in practical guidance and complicates efforts by CI stakeholders to adopt consistent and effective security practices.

To address this gap, we conducted a multivocal literature review of existing software supply chain security frameworks, government regulations, and academic studies.
We synthesized security practices from these diverse sources and developed a comprehensive analytical approach based on the \textbf{"4W+1H"}\cite{5ws_software_supply_chain} dimensions: \textbf{What, When, Where, Who, and How}, to systematically examine how software supply chain security is addressed across different contexts.
Specifically, we identified ten security categories (What), such as data protection, traceability, secure development, and incident response; mapped them across five software supply chain phases (When) and existing frameworks (Where); and associated them with three key stakeholder roles (Who): producers, operators, and consumers.
Recognizing the varying depth of implementation (How), we further classified each control into three levels of rigor, mandatory, recommended, and advanced, to support both baseline compliance and progressive maturity in practice.

Building on this foundation, we designed a structured, multi-layered security risk checklist that provides organisations with a practical approach for evaluating and improving their software supply-chain security posture.
The checklist, comprising 80 questions, is organised across life-cycle phases, maturity levels, and responsible roles, enabling targeted and context-aware assessments that align with real-world operational practices.

\section{Background}

\subsection{Challenges in Critical Infrastructure Sectors Implementation}
Despite growing recognition of the importance of software supply chain security, real-world implementation within CI sectors remains challenging.

\textbf{Fragmented Supply Chains and Limited Visibility:} CI operators often rely on software and hardware from dozens of suppliers, including second- and third-tier vendors, which complicates end-to-end assurance \cite{dnv2025supplychain}. 

\textbf{Resource and Skill Gaps:} Security teams are frequently focused on operational incident response, leaving little capacity for sustained, long-term supply chain security efforts \cite{isc22024workforce}.

\textbf{Misalignment Between Frameworks and CI Practices
Generic guidance:} Most frameworks are not tailored to sector-specific constraints, such as real-time system control or regulatory safety requirements.
Bridging the gap between framework design and the operational realities of CI sectors remains a central challenge in enhancing national cyber resilience \cite{tamanna2024analyzing}.



\subsection{Australia's Critical Infrastructure}



Australia's critical infrastructure sectors are increasingly vulnerable to software security risks, which pose significant threats to national security, economic stability, and public safety.

The 2023 Critical Infrastructure Resilience Strategy\footnote{Critical Infrastructure Resilience Strategy 2023. \url{https://www.cisc.gov.au/resources-subsite/Documents/critical-infrastructure-resilience-strategy-2023.pdf}} defines critical infrastructure as:
\begin{itemize}
    \item \emph{"...those physical facilities, supply chains, information technologies and communication networks, which if destroyed, degraded or rendered unavailable for an extended period, would significantly impact the social or economic wellbeing of the nation, or affect Australia's ability to conduct national defence and ensure national security."}
\end{itemize}

Australia's critical infrastructure spans a wide range (11 CI sectors) of essential sectors, \textit{Communications, Data storage/processing, Financial services/markets, Water and sewerage, Energy sector, Health care and medical, Higher education and research, Food and grocery, Transport, Space technology, Defence industry}, all of which are becoming increasingly dependent on digital technologies. Their resilience now relies heavily on the security and integrity of the software systems that keep them running. Safeguarding these systems has become a national priority under the Security of Critical Infrastructure Act 2018\footnote{Security CI Act 2018. \url{https://www.legislation.gov.au/C2018A00029/latest/versions}}.
From telecommunications (e.g., Telstra and Optus) to  financial services and market (e.g., Commonwealth Bank of Australia), water management (e.g., Sydney Water) and power grids (e.g., AGL Energy), software underpins almost every operational process. A single vulnerability or compromised update can cause outages, data breaches, or even public safety risks. Sectors such as healthcare (e.g., Australian Healthcare), education (e.g., CSIRO), and defence face particularly high stakes, as disruptions here can endanger lives, compromise sensitive data, or weaken national security.
Across all these sectors, the software supply chain plays a pivotal role in maintaining operational integrity and resilience. From firmware and embedded systems to cloud platforms and third-party libraries, the risks introduced by unverified or malicious software components must be systematically addressed. Establishing robust software supply chain security practices, including provenance tracking, vulnerability management, continuous monitoring, and supplier vetting, is essential to safeguarding Australia's critical infrastructure from evolving software security risks.

To bridge these gaps, it is crucial to align software security frameworks more closely with the specific, real-world contexts of critical infrastructure. 
In this paper, we examined how existing frameworks perform within particular critical infrastructure sectors. We proposed a comprehensive checklist that consolidates key elements from major frameworks, designed to provide practitioners with clearer guidance and to facilitate more effective risk management and compliance.


\section{Methodology}

This study employs a multivocal literature review (MLR) approach to comprehensively survey both industry and academic sources related to software supply chain security in critical infrastructure sectors. The MLR enables inclusion of diverse perspectives by combining authoritative industry frameworks, government regulations with peer-reviewed academic literature .

\subsection{Research Question}

Our paper is guided by the following research questions (RQs):



\begin{itemize}
    \item \textbf{RQ1:} What frameworks address software supply chain security practices in critical infrastructure sectors?
    \item \textbf{RQ2:} What specific security practices are proposed in existing frameworks?
    \begin{itemize}
        \item \textbf{RQ2.1:} How can the security requirements/practices identified in existing frameworks be systematically categorized (what)?
       \item \textbf{RQ2.2:} How can software security requirements/practices be implemented across the supply chain in terms of when, where, who, and how?
    \end{itemize}
\end{itemize}

    

\subsection{Study Selection}

We conducted the study selection through Google Search using the keyword query:
\emph{"(Software Supply Chain Security) AND (Framework OR Guidance OR Regulation OR Standard OR Best Practice) AND ((Sector-Specific) OR Industry)"}.
This search focused on established industry frameworks, government regulations, and relevant academic literature to identify comprehensive software security requirements for critical infrastructure sectors. The placeholder \emph{"(Sector-Specific)"} was systematically replaced with each CI sector keyword (e.g., \emph{"(ICT or Communications-specific)"}).
The selection criteria for including frameworks and literature were as follows:

\begin{itemize}
    \item \textbf{Widely adopted industry frameworks:} We focused on frameworks that have demonstrated significant adoption and influence across multiple sectors. These frameworks are typically developed by recognized standards organisations or government agencies and provide foundational guidance on securing software supply chains. Their inclusion ensures that our analysis captures broadly accepted best practices and requirements that shape industry security postures.

    \item \textbf{Sector-specific frameworks:} Considering that different critical infrastructure sectors face unique security challenges and regulatory requirements, we included frameworks tailored to particular domains. These sector-specific frameworks address specialized security risks and compliance needs, providing focused guidance for industry customers, e.g., communication, healthcare, financial and energy sector. 

    \item \textbf{Australian frameworks and regulations:} Given the applicable user case focus and regulatory context of our study, we prioritized Australian government frameworks and legislation. Including these regulations ensure our study reflects the local regulatory environment and aligns with national security priorities.

    \item \textbf{Academic literature:} To complement industry and regulatory sources, we included academic research and whitepapers that analyze software supply chain security risks and mitigation strategies in critical infrastructure from Google Scholar. Academic studies often provide deeper theoretical insights, empirical data, and innovative approaches that may not yet be codified in formal frameworks. Including these sources broadens the scope of our review and helps identify emerging challenges and best practices not fully covered by existing standards.
\end{itemize}

\subsection{Data Extraction}

From the selected frameworks and academic studies, we systematically extracted relevant information pertaining to software supply chain security requirements, categories, supply chain phases, and stakeholder roles. This process involved reviewing framework documentation, regulatory texts, and research articles to identify explicit security requirements, controls, and recommended practices. Key data elements were coded and organized into a structured format (see Table~\ref{tab:data_extraction}) to facilitate comparative analysis across sources. This structured extraction allowed us to synthesize and map diverse requirements and recommendations into coherent categories aligned with our research questions.

\begin{table}[htbp]
\centering
\small
\caption{Data extraction items}
\label{tab:data_extraction}
\begin{tabular}{p{0.13\textwidth}p{0.70\textwidth}p{0.05\textwidth}}
\toprule
\textbf{Data Item} & \textbf{Description}& RQ \\
\midrule
Framework & Name of the framework, regulation, or academic study.&RQ1 \\

Source Type & Classification as industry framework, government regulation, or academic literature.&RQ1 \\

Target CI Sector & The specific critical infrastructure sector(s) addressed (e.g., general, energy, healthcare).&RQ1 \\

Controls & Supply chain security recommendedation, practices or security requirements.&RQ2 \\

Categories & Classification of security requirements into broader categories or themes.&RQ2.1 \\


Phases & When practices are applied across the supply chain&RQ2.2  \\

Stakeholders & Who is responsible for their implementation (e.g., developers, managers). &RQ2.2 \\

Levels & How practices are addressed across different organisational or system layers&RQ2.2  \\

\bottomrule
\end{tabular}
\end{table}

\subsection{Data Analysis}

We systematically extracted and analyzed security requirements, controls, recommendations, and best practices, along with their corresponding categories, supply chain phases, and stakeholder roles from the selected sources. This comprehensive analysis directly addresses our research questions by enabling us to:

\begin{itemize}
    \item Review \textbf{what existing frameworks} address software supply chain security risks and proposed security practices in different critical infrastructure sectors.
    \item Categorize \textbf{software supply chain requirement} from  these frameworks focus on and develop a taxonomy of security practice categories.
    \item Determine at \textbf{which phases} of the software supply chain these risks should be mitigated and \textbf{who} should care about and implement software supply chain security controls.
    \item Understand \textbf{how risks are addressed} across different levels of — mandatory, recommended, and advanced measures — and reflect this in the structure of our assessment framework.
\end{itemize}

By integrating findings from multiple frameworks and research sources, our analysis provides a comprehensive understanding of software supply chain security requirements and practices. 
Based on these insights, we provided a structured and muti-layered set of checklist questions for security assessment and verification aligned with the identified software supply chain security categories, phases, and stakeholder responsibilities. 
This checklist supports targeted evaluation of software supply chain security posture and aids in the prioritization of risk mitigation activities across diverse sectors.









\section{Results} 

\subsection{RQ1: What existing frameworks address software supply chain security practices in CI sectors?}

\begin{table}[htbp]
\centering
\small
\caption{Software supply chain frameworks in CI }
\label{tab:frameworks}
\begin{tabular}{p{2.8cm} p{6.8cm} p{1.7cm} p{4cm}}
\toprule
\textbf{Reference / URL} & \textbf{Description} & \textbf{CI Sector} & \textbf{Focus} \\
\midrule
\href{https://www.legislation.gov.au/Details/C2018A00123}{\textbf{Sec. CI Act (2018)}} & Security of Critical Infrastructure Act 2018 — Australian legislation enhancing security and resilience of critical infrastructure through defined obligations for CI owners and operators. & All & Foundational Australian legislation defining security obligations for CI operators and owners. \\

\href{https://www.cyber.gov.au/resources-business-and-government/essential-cyber-security/ism}{\textbf{Australian ISM (2023)}} & Australian Government Information Security Manual — cybersecurity framework to protect IT and OT systems, applications, and data against cyber threats. & General & Widely used in Australian government cybersecurity governance. \\

\href{https://www.education.sa.gov.au/policies/shared/ict-cyber-security-standard.pdf}{\textbf{ICT CS (2022)}} & ICT Cyber Security Standard — defines cybersecurity controls for Department for Education IT systems; aligned with South Australian Cyber Security Framework (SACSF). & ICT/ Education & Focused on cybersecurity for education and communication infrastructure. \\

\href{https://csrc.nist.gov/pubs/sp/800/218/final}{\textbf{NIST SSDF v1.1 (2022)}} & Secure Software Development Framework (SSDF) v1.1 — U.S. federal guidance on secure software development practices to reduce vulnerabilities in software supply chains. & General & Widely recognized framework guiding secure software development best practices. \\

\href{https://www.cisa.gov/sites/default/files/publications/ESF_SECURING_THE_SOFTWARE_SUPPLY_CHAIN_DEVELOPERS.PDF}{\textbf{CISA (2023)}} & CISA Software Supply Chain Security Framework (Part 1 for Developers) — provides security guidelines and best practices for software supply chain risk management. & General & Practical advice and alerts on emerging threats to software supply chains. \\

\href{https://slsa.dev/}{\textbf{SLSA v1.0 (2021)}} & supply chain Levels for Software Artifacts — framework defining graduated levels of software supply chain security assurance to mitigate tampering and vulnerabilities. & General & Industry-recognized model for artifact integrity and provenance assurance. \\

\href{https://theupdateframework.io/}{\textbf{TUF (2019)}} & The Update Framework — open standard securing software update mechanisms to prevent malicious software supply chain attacks. & General & Provides robust metadata and trust delegation for secure updates. \\

\href{https://owaspsamm.org/model/}{\textbf{OWASP SAMM v2.1 (2022)}} & OWASP Software Assurance Maturity Model — maturity model for assessing and improving software security posture across development and acquisition; risk-driven and process-agnostic. & General & Maturity model for software security assurance. \\

\href{https://owasp.org/www-project-application-security-verification-standard/}{\textbf{OWASP ASVS v4.0.3 (2023)}} & OWASP Application Security Verification Standard — open standard for verifying security controls in web applications (e.g., XSS, SQL injection); customizable per organisation. & General & Application-level security assurance standard. \\

\href{https://cloudsecurityalliance.org/star}{\textbf{CSA STAR (2021)}} & Cloud Security Alliance Security, Trust, Assurance, and Risk (STAR) Registry — public registry documenting security and privacy controls for cloud offerings. & Financial & Cloud security and assurance transparency framework. \\

\href{https://www.nerc.com/pa/Stand/Pages/CIPStandards.aspx}{\textbf{NERC CIP-013 (2018)}} & North American Electric Reliability Corporation Critical Infrastructure Protection standards — regulatory standards for protecting bulk electric systems from cyber and physical threats, including supply chain risk management. & Energy & Regulatory standard focused on energy sector risk management. \\

\href{https://www.fda.gov/media/119933/download}{\textbf{MDCF (FDA, 2023)}} & Medical Devices Cybersecurity Framework — FDA guidance on cybersecurity considerations and information for medical devices, including premarket submissions. & Health & Medical device security framework ensuring premarket and postmarket assurance. \\[4pt]
\bottomrule
\end{tabular}
\end{table}

We investigated a range of widely recognized software supply chain security frameworks relevant to critical infrastructure sectors. Table~\ref{tab:frameworks} and Table~\ref{tab:academic_lit} provide an overview of these frameworks and related academic studies, including legislative acts, industry standards, practical guidance, and maturity models.

\textbf{Australian frameworks and regulations.}  
Frameworks such as the Security of Critical Infrastructure Act (Security CI Act) and the Australian Government Information Security Manual (ISM) are widely applied within the Australian public sector. Together, they establish the legislative and policy foundations for protecting IT and OT systems, defining clear obligations for operators of essential services and enhancing the overall resilience of CI sectors.

\textbf{General frameworks.}  
Internationally, frameworks such as the Secure Software Development Framework (SSDF), the supply chain Levels for Software Artifacts (SLSA), and the Update Framework (TUF) provide systematic guidance for developing and maintaining secure software supply chains. SSDF promotes secure development practices, SLSA introduces graduated assurance levels to verify the integrity and provenance of software artifacts, and TUF focuses on protecting the software update process from tampering or compromise.  
Complementing these, the Software Assurance Maturity Model (SAMM) and the Application Security Verification Standard (ASVS) offer structured methods for assessing and improving software security maturity and for verifying application-level controls such as authentication and input validation. In general, these frameworks provide a broad and technology-agnostic foundation for improving supply chain assurance across different CI sectors.

\textbf{Sector-specific frameworks.}  
Several frameworks focus on domain-specific needs. The Medical Devices Cybersecurity Framework (MDCF) issued by the U.S. FDA offers detailed guidance for securing medical devices, integrating cybersecurity into safety-critical design and regulatory review processes. The ICT Cyber Security Standard, aligned with the South Australian Cyber Security Framework, targets the education sector by defining controls to protect digital infrastructure in schools and universities.  
The North American Electric Reliability Corporation's Critical Infrastructure Protection (NERC CIP) standards establish regulatory requirements for safeguarding bulk electric systems against cyber and physical threats, including supply chain risk management. Similarly, the Cloud Security Alliance's STAR registry (CSA STAR) enhances transparency and assurance for cloud service providers, which is particularly relevant to financial and service-oriented sectors. These domain-specific frameworks highlight the importance of contextualized security measures that address the distinctive operational and regulatory characteristics of each sector.

Although these frameworks are comprehensive, few fully address the operational realities of specific CI domains. Most focus on general best practices and governance principles, leaving a gap in detailed, sector-oriented guidance for implementing and maintaining software supply chain security controls.

\textbf{Academic findings.}  
In parallel with industry efforts, a range of academic studies (see Table~\ref{tab:academic_lit}) examine how software supply chain security can be strengthened in CI contexts. The selected literature~\cite{paper1,paper2,paper3,paper1,paper5,paper6,paper7,paper8} highlights several recurring challenges, including the growing risks from industrial IoT devices, the introduction of new vulnerabilities through AI-enabled systems, and the persistent difficulty of managing software assurance across interdependent CI sectors. Recent work~\cite{paper8} further emphasizes the role of DevSecOps and adaptive risk management approaches in aligning security assurance with the unique operational environments of critical infrastructure.

\begin{table}[htbp]
\centering
\small
\caption{Selected academic literature}
\label{tab:academic_lit}
\begin{tabular}{p{0.8cm} p{6cm} p{8cm}}
\toprule
\textbf{Ref} & \textbf{Title} & \textbf{Focus} \\
\midrule
\cite{paper1} & \textit{Understanding the Challenge of Cybersecurity in Critical Infrastructure Sectors} (2019) & Overview of broad cybersecurity challenges and risk factors affecting various CI sectors; highlights systemic vulnerabilities and interdependency risks. \\

\cite{paper2} & \textit{Critical Infrastructures: IT Security and Threats from Private Sector Ownership} (2018) & Analysis of security risks arising from privatization and IT outsourcing in CI environments; identifies governance and accountability gaps. \\

\cite{paper3} & \textit{Cybersecurity Considerations for Industrial IoT in Critical Infrastructure Sector} (2020) & Examination of IoT-related vulnerabilities and recommended protections in industrial CI settings, with emphasis on OT system security. \\

\cite{paper4} & \textit{Critical Infrastructure and Cyber Security} (2020) & Survey of sector-specific cybersecurity challenges and potential regulatory responses across multiple CI domains. \\

\cite{paper5} & \textit{Critical Infrastructure Protection: Generative AI, Challenges, and Opportunities} (2024) & Exploration of AI-related risks and opportunities for enhancing critical infrastructure cybersecurity and resilience. \\

\cite{paper6} & \textit{Contemporary Cyber Threats to Critical Infrastructures: Management and Countermeasures} (2022) & Detailed discussion on emerging cyber threats and adaptive defense mechanisms for CI resilience and protection. \\

\cite{paper7} & \textit{Recommendations for Effective Security Assurance of Software-Dependent Systems} (2021) & Compilation of best practices and assurance techniques for software security in CI systems, emphasizing verification and validation. \\

\cite{paper8} & \textit{Towards a DevSecOps-Enabled Framework for Risk Management of Critical Infrastructures} (2023) & Proposal of a DevSecOps-driven risk management framework tailored to CI software security assurance and continuous governance. \\
\bottomrule
\end{tabular}
\end{table}

\textbf{Findings (RQ1).}  
Overall, we found that existing general sector frameworks, including SSDF, SLSA, TUF, SAMM, and ASVS, provide a solid foundation for software supply chain security but remain largely generic and fragmented. 
Their broad coverage and varying levels of abstraction often create implementation challenges , particularly for CI operators who must comply with multiple overlapping standards. Furthermore, most frameworks are not sufficiently aligned with the operational and regulatory realities of CI sectors, where resource constraints, multi-tier supply chains, and real-time control requirements introduce additional complexity.  
While the current landscape offers a strong starting point for software supply chain security assurance, its effectiveness in CI environments is limited by both framework fragmentation and contextual misalignment. Our findings highlight the need for harmonized and sector-aware frameworks that better integrate security requirement and practices into the practical operations of critical infrastructure.




\subsection{RQ2: What specific security practices are proposed in existing frameworks?}


We answered RQ2 in terms of \textbf{"4W+1H"} approach, \textbf{What}, \textbf{When}, \textbf{Where}, \textbf{Who}, and \textbf{How}, to systematically examine software supply chain security practices across existing frameworks.






\begin{figure}[!htbp]
    \centering
    \begin{forest}
    for tree={
        anchor=west,
        grow'=east,
        rectangle,
        edge path={
            \noexpand\path[\forestoption{edge}]
              (!u.parent anchor) -- +(1em,0) |- (.child anchor)\forestoption{edge label};
        },
        draw,
        parent anchor=east,
        child anchor=west,
        rounded corners,
        top color=white,
        bottom color=gray!20,
        edge={thick},
        l sep=15pt,
        s sep=2pt,
        font=\small,
    },
    for level={0}{text width=0.20\textwidth},
    for level={1}{text width=0.30\textwidth},
    for level={2}{text width=0.30\textwidth},
    [Software supply chain practice for CI
        [Accountability framework
            [Role and responsibility]
        ]
        [Data protection
            [Access control]
            [Data definition]
            [Data policy]
        ]
        [Incident management
            [Incident report]
            [Incident response]
            [User report]
        ]
        [Risk management
            [Risk identification]
            [Risk program]
            [Risk report]
            [Risk assessment]
            [Risk control]
            [Logging and monitoring]
        ]
        [Secure environment
            [Environment protection]
            [Repository system]
            [Security check criteria]
            [Software development security policy]
            [Development model]
        ]
        [Secure software development
            [Reuse of existing components]
            [Secure coding]
            [Design review]
            [Security testing]
        ]
        [Software build and deployment
            [Build platform]
            [Build process]
            [Build verification]
            [Deployment verification]
        ]
        [Software traceability
            [Component registration]
            [Data archiving]
            [Provenance management]
            [Record keeping]
            [Stakeholder communication]
        ]
        [Software sourcing and procurement
            [Product evaluation]
            [Security requirement]
            [Source verification]
            [Contract management]
        ]
        [Software update
            [Attack detection]
            [Update verification]
        ]
    ]
    \end{forest}
    \caption{Software supply chain practice categories}
    \label{fig:ssc_practice_ci_tree}
\end{figure}
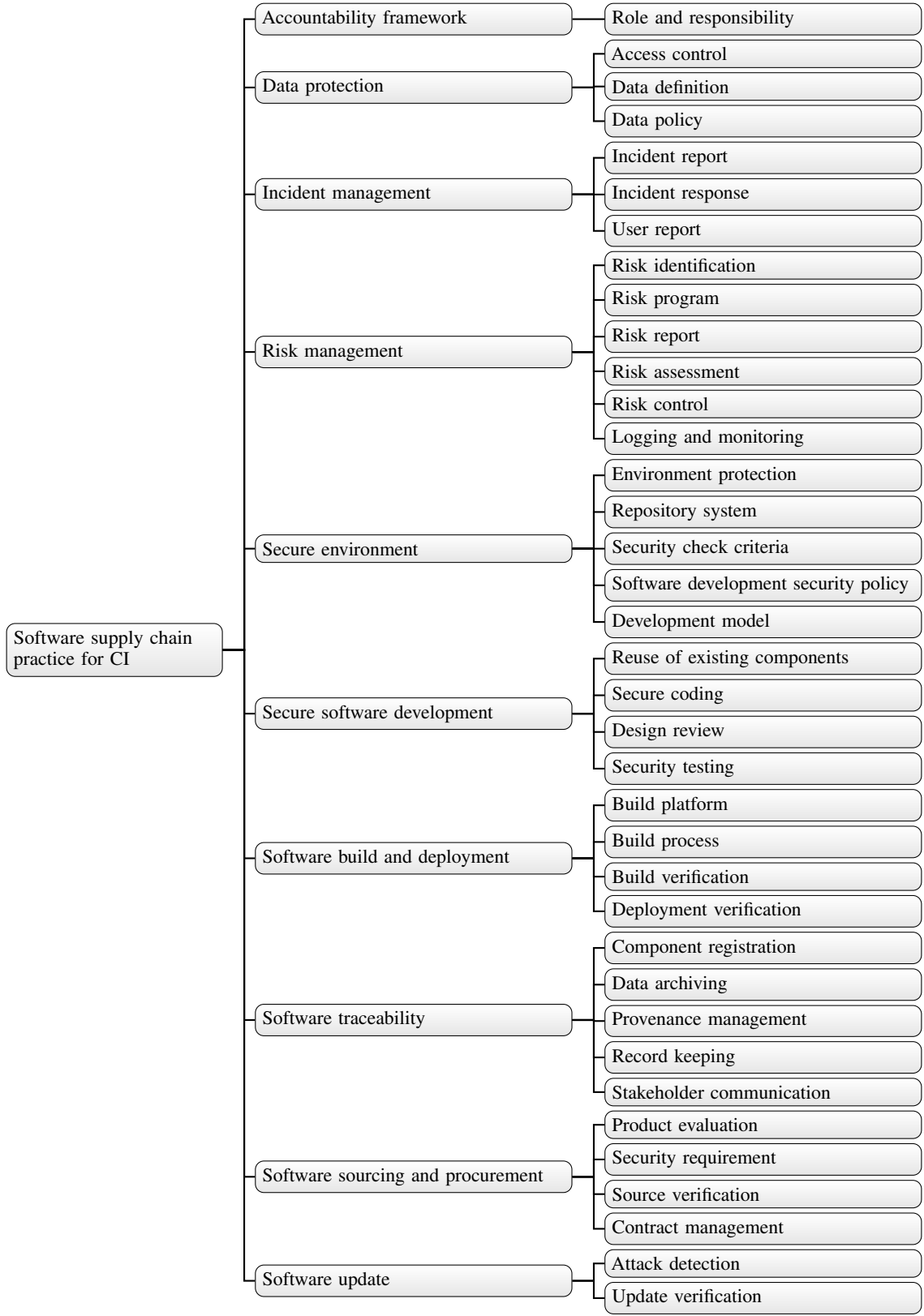

\subsubsection{RQ2.1: How can the security requirements and practices identified in existing frameworks be systematically categorized? (what)} 

To understand how existing frameworks address software supply chain security, we analyze \textbf{What} (by category, what specific practices are applied across the supply chain).
We extracted and analyzed security practices from RQ1's identified widely adopted standards, regulations, and maturity models. Through this process, we identified ten core security categories and associated sub-categories that reflect recurring concerns and control areas across these frameworks. These categories span both technical and governance domains and serve as the foundation for understanding the scope and focus of current software supply chain guidance.
Figure~\ref{fig:ssc_practice_ci_tree} presents 10 categories and 35 associated sub-categories, providing a high-level view of how software supply chain security practices are organized across software supply chain existing frameworks. 

\begin{sidewaysfigure} [htbp]
    \centering
    \includegraphics[width=\textwidth]{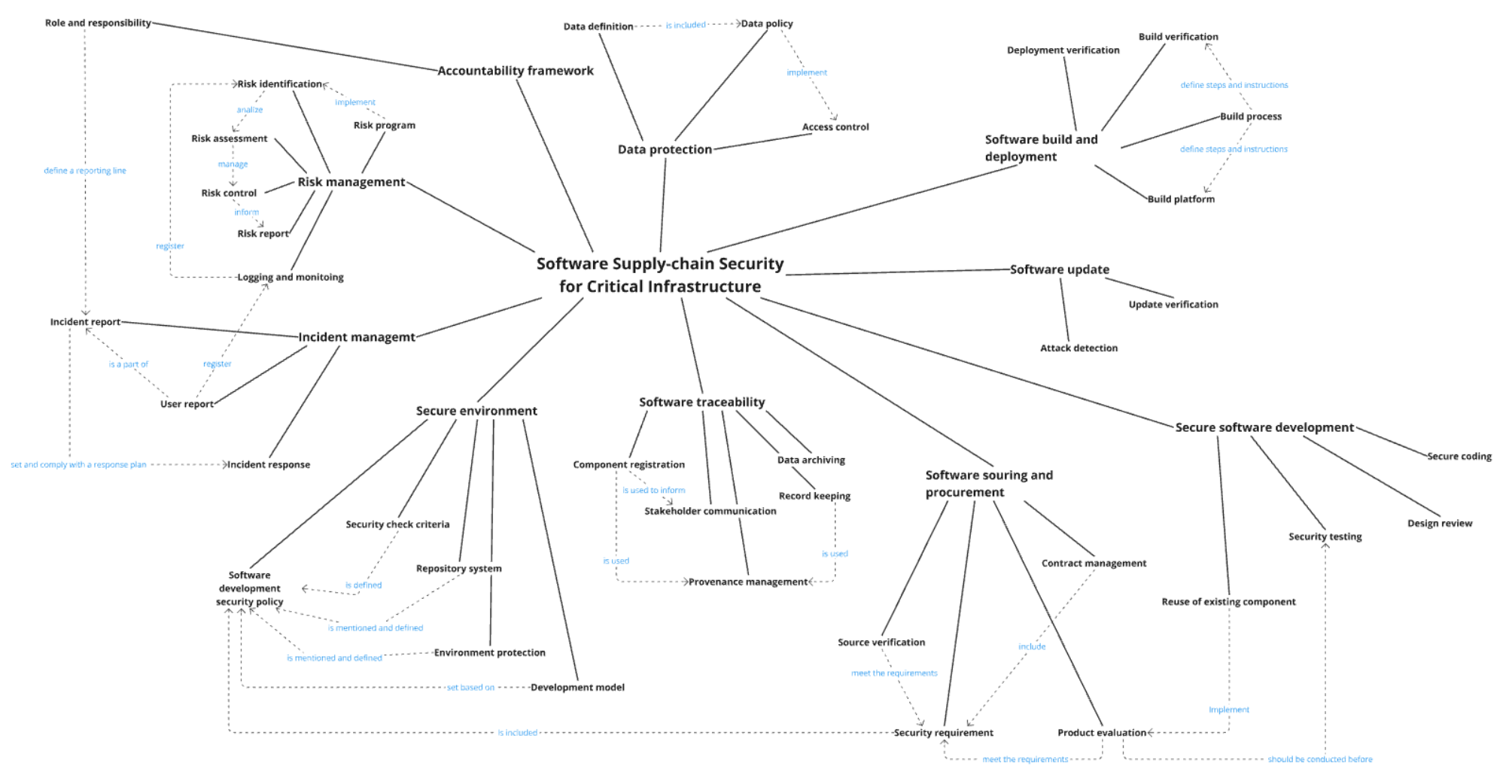}
    \caption{A concept map of software security practices.}
    \label{fig:Concept map}
\end{sidewaysfigure}

A visual concept map illustrating the relationships among software security practices is shown in Figure~\ref{fig:Concept map}.

\textbf{What: By Category}

\textbf{Accountability framework} emphasizes the need for clearly defined roles, responsibilities, and lines of communication across stakeholders. This ensures that every actor involved—from infrastructure owners to update signers—understands their obligations, which helps reduce risks of miscommunication, delays, or vulnerabilities due to unclear ownership.
We identified one key sub-category under the accountability framework: role and responsibility. This covers internal documentation, CI sector coordination, and the definition of cryptographic signing roles (e.g., root, snapshot, timestamp) in update mechanisms.

\textbf{Data protection} involves establishing formal policies and processes to safeguard information throughout the organisation. This includes the classification and definition of data based on sensitivity, restricting data usage and disclosure in line with data policies, and implementing access controls to ensure that only authorized personnel and systems can access sensitive data such as source code or configuration files. These measures are critical for maintaining the integrity and confidentiality of software assets within critical infrastructure sectors. 
We identified three sub-categories within data protection: data policy, data definition, and access control.

\textbf{Incident management} focuses on an organisation's ability to report, respond to, and learn from cybersecurity incidents. It involves having a structured reporting system, a clear classification of incidents, and regularly reviewed response plans to address threats effectively and minimize impact.

Three sub-categories fall under this category: incident report– systems and processes for reporting incidents to relevant stakeholders, incident response– plans for responding to and recovering from incidents, along with compliance and review mechanisms, and user report– mechanisms that allow users to report bugs, anomalies, or suspicious activities, enhancing visibility and early detection.

\textbf{Risk management} is a structured approach to identifying, assessing, mitigating, and monitoring risks that could compromise system integrity or availability. This includes establishing a formal risk management program, generating regular risk reports, identifying and evaluating direct or indirect risks (such as cyberattacks), and implementing mitigation or backup plans.
There are six sub-categories within risk management: risk program, risk identification, risk assessment, risk report, risk control, and logging and monitoring. Each addresses different aspects of maintaining operational and cyber resilience in CI sectors.

\textbf{Secure environment} is essential to ensuring software development is resilient to threats across all phases of the software supply chain. This includes not only enforcing security policies and criteria but also protecting the physical and logical environments where software is developed, built, tested, and distributed.
We identified six sub-categories to address secure environment: software development security policy, security check criteria, environment protection, repository system, and development model.
These practices collectively help maintain a hardened and monitored environment that minimizes vulnerabilities introduced during development and delivery.

\textbf{Secure software development} focuses on embedding security throughout the entire software development life cycle, from design and component reuse to coding practices and security testing. The goal is to proactively identify, prevent, and mitigate vulnerabilities and flaws before deployment.
There are four sub-categories: design review, reuse of existing component, secure coding and security testing.
These practices aim to reduce attack surfaces early in development and ensure that security is a continuous and integrated part of the development process.

\textbf{Software build and deployment} is to ensure that software is built, signed, and deployed in a controlled, verifiable, and tamper-resistant environment. It involves securing build processes, validating provenance, and continuously verifying the integrity of deployed artifacts.
We identified four sub-categories to ensure that both the build and deployment stages are resistant to tampering, misconfiguration, and supply chain compromise: build platform, build process, build verification, and deployment verification. 
These practices are primarily derived from the supply chain Levels for Software Artifacts (SLSA), but also draw on the Secure Software Development Framework (SSDF), the OWASP Application Security Verification Standard (ASVS), and other relevant frameworks.

\textbf{Software traceability} is essential for ensuring transparency, integrity, and accountability throughout the software supply chain. It involves tracking and documenting components, dependencies, and updates to establish clear provenance and enable verification of authenticity.
Five sub-categories are identified for software traceability: component registration, stakeholder communication, provenance management, record keeping, data archiving.
Component registration refers to the documentation and sharing of software components (e.g., SBOM).
Stakeholder communication involves sharing traceability data with relevant parties to ensure transparency.
Provenance management focuses on tracing software artifacts back to their source and build processes.
Record keeping is essential for maintaining logs, metadata, and update records to support continuous monitoring.
Data archiving ensures the retention of verification and provenance data for future reference.

\textbf{Software souring and procurement} focuses on ensuring that acquired software meets security standards through well-defined requirements, thorough evaluation, formal contracts, and integrity verification of source components.
Four sub-categories are identified to help prevent vulnerabilities introduced through third-party software and strengthen trust in procured or externally sourced components: security requirement, product evaluation, contract management, and source verification.

\textbf{Software update} is to ensure integrity, authenticity, and resilience against tampering or attacks during the update process. This includes verifying metadata, validating cryptographic signatures, and detecting various types of update-related attacks.
We identified two sub-categories within software update: update verification and attack detection.
While "update verification" is to ensure that update metadata is properly signed, validated, and untampered, supporting authenticity through cryptographic checks and defined roles, "attack detection" focuses on identifying rollback attacks, freeze attacks, and compromised signing keys, enabling organisations to detect and respond to suspicious behaviors during the update process.

\subsubsection{RQ2.2: How can software security requirements and practices be implemented across the supply chain in terms of when, where, who, and how?}




Given the complex and distributed nature of software supply chain activities, software security practices are implemented across following interrelated dimensions: \textbf{When} (by phase, when practices are applied across the supply chain), \textbf{Where} (by framework, which practices are covered in which existing framework), \textbf{Who} (by stakeholder, who is responsible for their implementation), and \textbf{How} (by level, how practices are addressed across different organisational or system layers).

\begin{table}[htbp]
\centering
\small
\caption{Distribution of software supply chain security practices by life cycle phase}
\label{tab:category_phase}
\begin{tabular}{p{0.2\textwidth}*{5}{p{0.1\textwidth}}p{0.1\textwidth}}
\toprule
\textbf{Phase} & \textbf{GOV} & \textbf{DEV} & \textbf{DEP} & \textbf{PRC} & \textbf{TRC} & \textbf{Total} \\
\midrule
Preparation             & 4  & 6  & 0  & 1  & 1 & 13 \\
Procurement/acquisition & 0  & 0  & 0  & 9  & 0 & 9  \\
Development             & 6  & 14 & 0  & 0  & 2 & 18 \\
Deployment              & 0  & 4  & 11 & 0  & 2 & 17 \\
Post-deployment         & 5  & 4  & 2  & 0  & 5 & 16 \\
Entire life cycle       & 6  & 0  & 1  & 0  & 0 & 7  \\
\midrule
\textbf{Total}          & 21 & 28 & 14 & 10 & 10 & 80 \\
\bottomrule
\end{tabular}

\vspace{1ex}
\begin{flushleft}
\textbf{Legend:} \\
\textbf{GOV}: Governance and risk Management \quad
\textbf{DEV}: Secure environment and development \quad
\textbf{DEP}: Deployment and update \quad
\textbf{PRC}: Procurement \quad
\textbf{TRC}: Software traceability
\end{flushleft}
\end{table}
\textbf{When: By phase}

To support effective adoption and implementation of software supply chain security practices, it is essential to understand when (at which life cycle phase) and by whom (which stakeholder) each control should be applied. However, most existing frameworks either focus narrowly on a single phase (e.g., SLSA for deployment, TUF for post-deployment) or provide only high-level guidance without explicitly mapping responsibilities to specific actors (e.g., NIST SSDF, CISA guidance, and SAMM). In this section, we analyze how security practices are distributed across the software supply chain life cycle, spanning preparation, procurement/acquisition, development, deployment, and post-deployment.

Our analysis reveals that software security practices span the entire software supply chain, with different types of security risks emerging at distinct phases (Table~\ref{tab:category_phase}). 

The development phase is the most heavily covered, with 18 items, particularly concentrated in secure software development (12) and software build and deployment (2), reflecting the critical importance of secure coding practices, toolchain integrity, and testing. This is followed by the deployment phase (17), which is especially rich in controls related to software build and deployment (11) and software traceability (4), underscoring the emphasis on provenance, build verification, and artifact integrity at release time.

The post-deployment phase also shows strong representation (16 items), primarily in software update (6), incident management (3), and data protection (3), indicating a clear need for ongoing monitoring, patch validation, and response planning. The preparation phase (13) highlights the need for early-stage planning in areas such as risk management, incident management, and secure environment setup.

Interestingly, while fewer in total, the procurement/acquisition phase (9) reflects targeted concerns related to software sourcing and procurement, such as contract requirements and source verification. The entire life cycle category (7) includes frameworks that span all phases, emphasizing integrated, end-to-end approaches to security.

Overall, this mapping illustrates that security risks must be addressed continuously, from initial preparation and procurement to development, deployment, and ongoing operations. Each phase plays a unique role in maintaining the integrity and trustworthiness of software systems.

\textbf{Where: By framework}

The distribution of security requirement/control items across software supply chain shows that certain source frameworks prioritize specific stages of the software supply chain (Table~\ref{tab:category_framework}).

\begin{table}[htbp]
\centering
\small
\caption{Distribution of supply chain phases across frameworks}
\label{tab:category_framework}
\begin{tabular}{p{3cm} c c c c c c c c c c c}
\toprule
\textbf{Phase} & \textbf{SCI Act} & \textbf{SLSA} & \textbf{NIST} & \textbf{TUF} & \textbf{CISA} & \textbf{OWASP} & \textbf{ISM} & \textbf{MDCF} & \textbf{OWASP} & \textbf{ICT }  \\
\midrule
Preparation & 4 & 0 & 5 & 0 & 1 & 0 & 0 & 0 & 0 & 1  \\
Procurement/acquisition & 0 & 0 & 2 & 0 & 3 & 0 & 0 & 0 & 0 & 0  \\
Development & 2 & 0 & 13 & 0 & 0 & 2 & 1 & 1 & 1 & 0  \\
Deployment & 1 & 11 & 0 & 0 & 1 & 2 & 1 & 1 & 0 & 0  \\
Post-deployment & 2 & 0 & 1 & 11 & 1 & 1 & 0 & 1 & 0 & 1 \\
Entire life cycle & 3 & 0 & 1 & 0 & 1 & 0 & 0 & 0 & 0 & 0 \\
\hline
\textbf{Total} & \textbf{12} & \textbf{11} & \textbf{22} & \textbf{11} & \textbf{9} & \textbf{5} & \textbf{2} & \textbf{3} & \textbf{1} & \textbf{2}  \\
\bottomrule
\end{tabular}
\end{table}

NIST SSDF stands out as the most comprehensive source with heavy emphasis on the development phase and notable presence across all phases, including preparation, deployment, post-deployment. This demonstrates NIST's commitment to secure software development through extensive guidance on coding practices, toolchain validation, and environmental controls.

SLSA contributes exclusively to the deployment phase (11), reflecting its targeted focus on build verification, provenance, and artifact integrity during the release process. In contrast, TUF also concentrates on post-deployment (11), highlighting the importance of update validation and attack detection after software is released.

The Security CI Act contributes to several phases, with a strong focus on development (2), post-deployment (2), and entire life cycle (3), indicating its role in enforcing both early- and late-stage controls in critical infrastructure contexts. CISA, while contributing fewer total risk items, is particularly active in the deployment (6) and development (1) phases.

Other frameworks such as OWASP, ISM, ICT CS, and MDCF show more distributed, focusing on specific areas like secure coding, software verification, or risk management.

This breakdown shows how some frameworks provide end-to-end guidance, like NIST SSDF and the Security CI Act, while others focus on specific security-critical phases, like SLSA (deployment) or TUF (post-deployment). Yet, they form a diverse landscape of tools and practices tailored to different parts of the software supply chain.

\textbf{Who: By stakeholder}

\begin{table}[htbp]
\centering
\small
\caption{Distribution of software supply chain security practices by stakeholder}
\label{tab:category_stakeholder}
\begin{tabular}{p{2cm} c c c c c c c}
\toprule
\textbf{Stakeholder} & \textbf{Preparation} & \textbf{Procurement} & \textbf{Development} & \textbf{Deployment} & \textbf{Post-dep} & \textbf{Entire Life} & \textbf{Total} \\
\midrule
Consumer & 0 & 6 & 0 & 1 & 1 & 0 & 8 \\
Manager  & 4 & 0 & 0 & 0 & 3 & 2 & 9 \\
Producer & 5 & 1 & 15 & 11 & 10 & 2 & 44 \\
\hline
\textbf{Total} & \textbf{9} & \textbf{7} & \textbf{15} & \textbf{12} & \textbf{14} & \textbf{4} & \textbf{61} \\
\bottomrule
\end{tabular}
\end{table}

The analysis of stakeholder responsibilities across the software supply chain presents that the "operator" role dominates, accounting for 62 out of 80 total security requirement items (Table~\ref{tab:category_stakeholder}). 
A "producer" side role and refers to be represented by software developers, engineers, and IT professionals, responsible for designing, developing, and maintaining software systems.
This indicates that day-to-day operational responsibilities, such as secure implementation, verification, monitoring, and incident handling, largely fall to operators, especially during the development (15), deployment (12), and post-deployment (14) phases. In contrast, "managers" (is also a producer side role,  including executive leadership, IT/security senior management, and decision-makers) are more engaged in early-stage phases such as preparation (4) and post-deployment (3) and entire life-cycle (2), where policy setting, risk assessment, and contractual decisions are made.

Meanwhile, consumers (including external customers, internal employees, and other stakeholders who interact with software systems, impacted by the security of software products) while less frequently mentioned (8 in total), are mainly responsible for procurement/acquisition (6) where they must evaluate, verify, and ensure trust in sourced software and deployment (1) or post-deployment (1) in limited cases.

This analysis clearly shows a distributed responsibilities between stakeholders. While operators take the majority of implementation responsibility, managers lead planning and oversight, and consumers play a selective but important role in software trust and validation.

\textbf{How: By level}


\textbf{Level classification in existing framework}.
Software security frameworks vary in how explicitly they define levels of control or maturity levels for software security requirement. 
According to our survey, we identified a common need to tailor security controls to organisational maturity, risk tolerance, and system criticality. Yet, only a few frameworks provide clearly layered controls for users.

SLSA introduces three levels of supply chain assurance, rating from basic provenance tracking (Level 1) to strong integrity guarantees (Level 3). The following presents the details of the three levels of security for software deployment.
Each level builds upon the previous one, progressively strengthening the trustworthiness and verifiability of the software build process. 
At the foundational level (Level 1), SLSA requires a consistent and reproducible build process, along with the existence and distribution of provenance information that records the build environment, process, and inputs.
At the intermediate level (Level 2), additional requirements are introduced, including the use of a hosted build platform and digitally signed provenance, ensuring authenticity and traceability across the supply chain.
At the highest level (Level 3), SLSA enforces strong platform isolation to prevent cross-build contamination, and mandates unforgeable provenance, where cryptographic signing keys are securely managed and inaccessible to user-defined processes.



Similarly, SAMM defines maturity levels across security practices to guide incremental improvement.
At the initial level (Level 1), organisations begin establishing basic policies for governance, conduct limited threat modeling, and introduce secure coding guidelines, though implementation remains inconsistent.
At the intermediate level (Level 2), policies and processes become standardized and measurable, with formalized design reviews, threat modeling, and consistent application of secure coding standards across teams.
At the highest maturity level (Level 3), software security becomes an integral part of enterprise risk management, supported by automated tooling, verified design components, and continuous feedback loops for improvement.

Meanwhile, other frameworks such as NIST SSDF and CISA guidance have been widely used, but lack explicit layered implementation paths.

In summary, while frameworks like SLSA and SAMM define maturity levels or assurance tiers, our review identified notable inconsistencies in how control granularity and implementation depth are articulated across frameworks. Some frameworks, such as NIST SSDF or CISA guidance, offer best-practice recommendations but do not explicitly tier their controls based on organisational maturity, risk exposure, or resourcing capacity. As a result, CI sector organisations may struggle to determine which controls are essential, which are aspirational, and how to prioritize implementation under operational constraints.

\textbf{Level classification in our survey}. To address this gap and support context-aware adoption, we introduce a simplified three-level classification model. This model allows organisations to incrementally adopt supply chain security practices based on their existing capabilities and risk profile, without needing to fully implement an entire framework.
We defined three levels: \textbf{mandatory}, \textbf{recommended}, and \textbf{advanced} as follows.
The definitions were based on the layered security principles in frameworks such as SLSA and OWASP SAMM, and observed practice patterns across government, industry, and sector-specific guidelines. 
\begin{itemize}
    \item \textbf{Mandatory} level includes foundational requirements that are often prescribed by regulation or considered essential across most frameworks. Examples include maintaining a consistent build process, distributing provenance artifacts, and establishing basic security policies. These controls are designed to provide minimum assurance and are typically required to meet baseline compliance standards. In this study, we primarily classify security controls derived from government regulatory or standard sources (e.g., \textit{Security CI Act})under this level.
    \item \textbf{Recommended} level reflects enhanced practices that improve an organisation's security posture beyond the baseline. These include adopting hosted build platforms to prevent tampering, conducting formal threat modeling, or performing structured code reviews based on secure coding standards. While not always mandatory, these controls are widely endorsed in best-practice frameworks (e.g., NIST SSDF, CISA) and are particularly important for systems with moderate to high impact levels.

    \item \textbf{Advanced} level includes high-assurance techniques aimed at organisations managing sensitive or critical software supply chains. Examples include ensuring build isolation and provenance unforgeability, implementing automated security testing tools and secure design pattern libraries, and enforcing separation of duties and hardened CI/CD pipelines (e.g., TUF, SLSA, OWASP SAMM). These controls are resource-intensive but provide robust protection against sophisticated supply chain threats.
\end{itemize}

\section{Discussion}

Through this study, our survey highlights the need for future research and policy development to move beyond one-size-fits-all security frameworks. Based on the current landscape, there is a clear opportunity to develop sector-specific profiles of supply chain security practices that better reflect the operational realities and criticality of each CI domain.

Furthermore, we gained deeper insights into key security practices necessary for managing diverse security risks across software supply chain life cycle. While existing frameworks provide valuable guidance, many lack comprehensive, end-to-end coverage of the software supply chain. Some frameworks are phase-specific. For example, SLSA focuses primarily on the deployment phase, while TUF targets post-deployment integrity. Others, such as NIST SSDF, offer broader coverage across the software life cycle and stakeholder groups, yet often lack structured maturity levels that can guide organisations in progressively implementing controls based on risk, resource constraints, or operational context.

\subsection{Software Supply Chain Security Checklist}

To fill above gaps, we have developed a software security risk checklist based on findings from previous \textbf{"4W+1H"} results. This checklist consolidates key insights from our analysis of existing frameworks and classifications, offering a practical, adaptable tool for improving software supply chain security in CI sectors. It is structured to support phased adoption across multiple life cycle stages and stakeholder roles, and categorised according to three implementation levels: mandatory, recommended, and advanced, helping organisations align security efforts with their maturity and risk profiles.

\textbf{Overview structure of checklist.}
This checklist consolidates security controls/requirements/practices identified across leading frameworks, including the Security CI Act, SLSA, NIST SSDF, and others, into a standardised, actionable format.

\begin{figure} [htb]
    \centering
    \includegraphics[width=\textwidth]{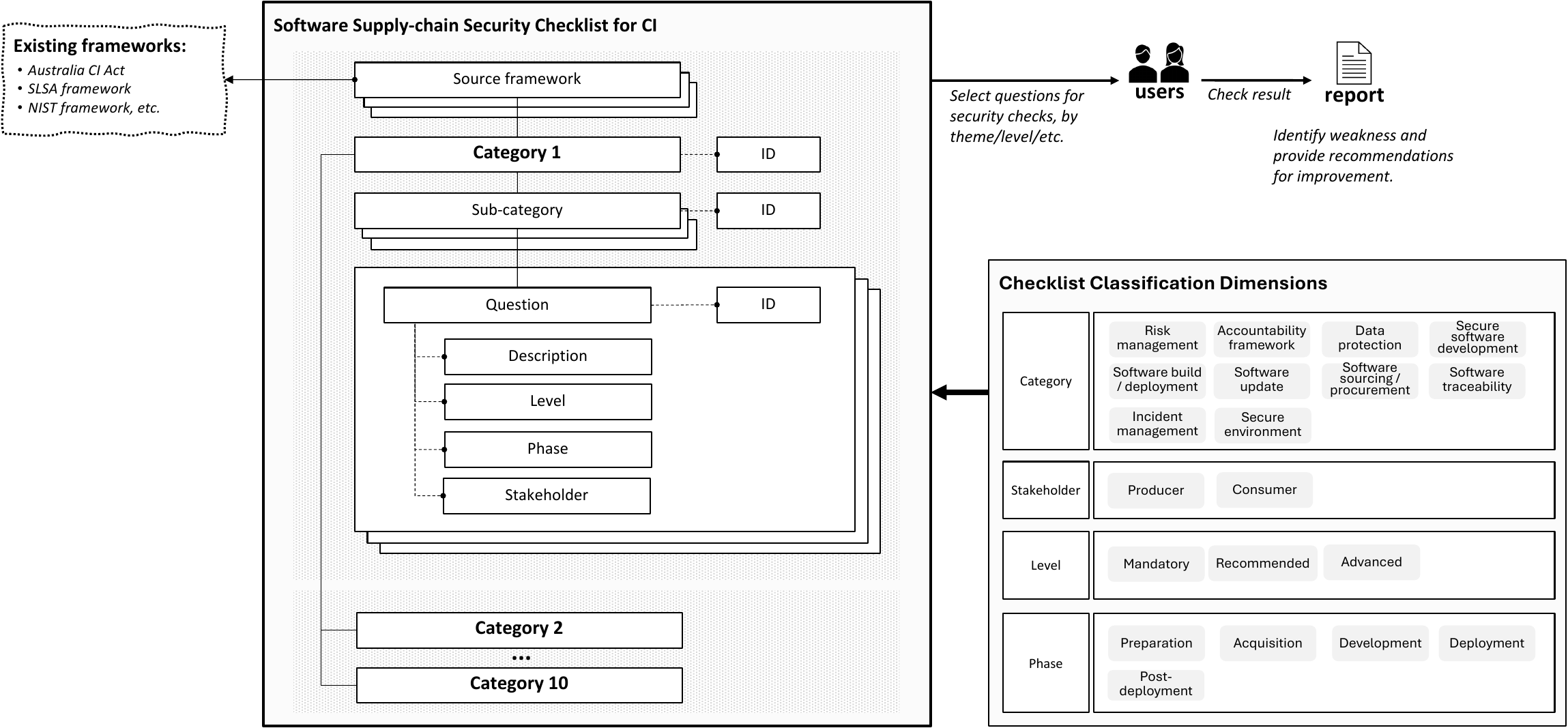}
    \caption{Overview of the checklist}
    \label{fig:checklist_overview}
\end{figure}

As illustrated in Figure \ref{fig:checklist_overview}, the checklist is hierarchically organised. Each entry is derived from a source framework and mapped into a category (e.g., risk management, software update) and sub-category (e.g., threat detection, access control). Each checklist item contains a question, description, implementation level, relevant software supply chain phase, and associated stakeholders (Table \ref{tab:checklist_example_traceability}). Some questions can be further divided into sub-questions based on more detailed requirements. The full checklist \footnote{Software supply chain checklist.\url{https://docs.google.com/spreadsheets/d/1EY4HT4bI05Ea2YNV0iV4TnOeOuJ1fKIM/edit?usp=sharing&ouid=102592420087316267740&rtpof=true&sd=true}} with detailed questions is available online.

To ensure consistent tagging and navigation, we defined four classification dimensions for each question in checklist as follows.

\begin{itemize}
    \item Question - Category: Aligns each question with a specific category of software security practice (e.g., software build and deployment, data protection, secure environment).
    \item Question - Stakeholder: Identifies whether the responsibility of this checklist question primarily lies with producers, consumers, or both.
    \item Question - Level: Organises questions into three implementation levels such as mandatory, recommended, and advanced. This is to support scalable adoption based on organisational risk and capacity.
    \item Question - Phase: Links each question to a supply chain phase in the software life cycle, including preparation, acquisition, development, deployment, and post-deployment.
\end{itemize}

This layered classification enables users to navigate and tailor the checklist according to their operational role, system maturity, and risk exposure. It also supports traceability back to the originating frameworks, ensuring transparency and adaptability across sectors.

\begin{figure} [htb]
    \centering
    \includegraphics[width=0.8\textwidth]{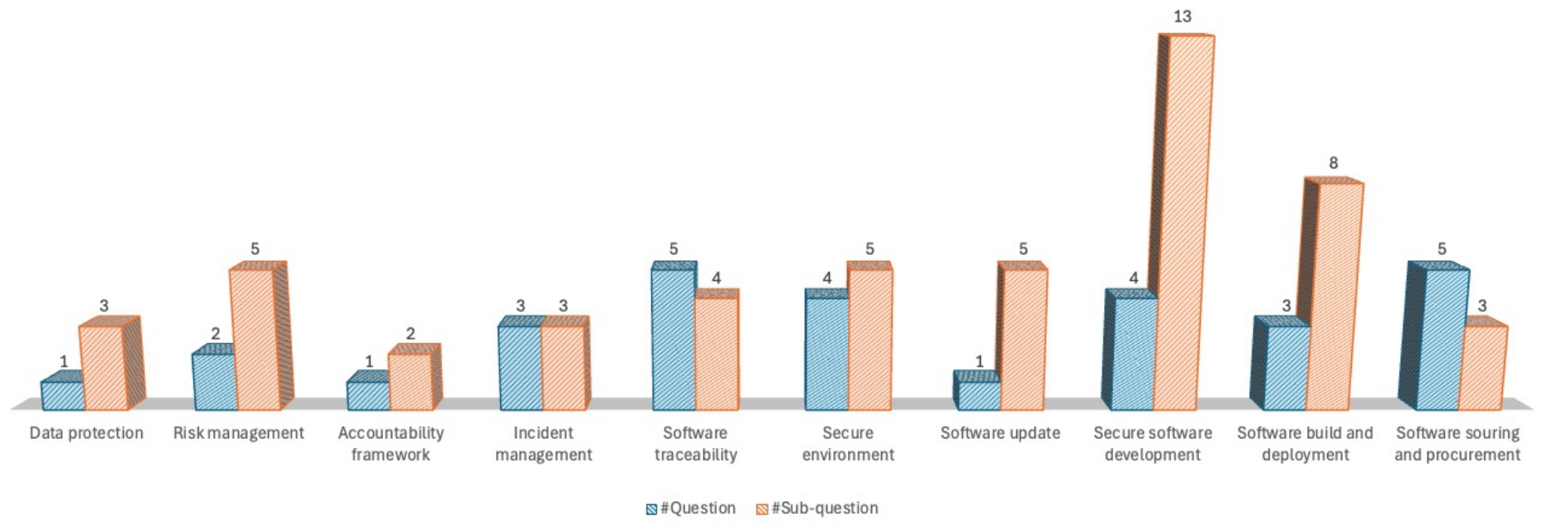}
    \caption{Checklist summary}
    \label{fig:checklist_summary}
\end{figure}

\textbf{Checklist question summary.} Table \ref{fig:checklist_summary} presents a summary of the checklist.
The category with the highest number of items is Secure software development, comprising 17 items (4 questions and 13 sub-questions), reflecting the complexity and centrality of development-phase controls in the supply chain. This is followed by Software build and deployment (11 items) and Secure environment and Software traceability (9 items each), indicating strong emphasis on securing both development infrastructure and traceability of artifacts. While Data protection and Accountability framework contain fewer items, their inclusion ensures foundational governance and compliance are not overlooked. Overall, the checklist comprises 29 top-level questions and 51 sub-questions, totaling 80 items across all categories.





\textbf{Example checklist question (Software traceability).} To illustrate the structure and use of our checklist, Table \ref{tab:checklist_example_traceability} provides an example focused on the Software traceability category. This category includes practices related to component registration, provenance, metadata tracking, and stakeholder communication. Each item is tagged with its source framework, implementation level, software life cycle phase, and responsible stakeholder group, supporting contextual adoption across CI sectors.

\begin{table}[htb]
\centering
\caption{Example checklist items from the category: Software traceability}
\resizebox{\textwidth}{!}{
\begin{tabular}{@{}l l l l l l@{}}
\toprule
\textbf{Source} & \textbf{Sub-category} & \textbf{Question (abridged)} & \textbf{Lvl} & \textbf{Phase} & \textbf{Stakeholder} \\
\midrule
Security CI Act & Component registration & Q:Process to register SBOM and attestations? & M & PR & P/M \\
Security CI Act & Stakeholder communication & Sub-Q:Share SBOM info with stakeholders? & M & D & P/O \\
SLSA & Provenance management & Q:Provenance to trace software back to source? & R & D & P/O \\
SLSA & Stakeholder communication & Q:Package ecosystem rule for provenance? & R & D & P/O \\
NIST SSDF & Record keeping & Q:Record keeping with tool support? & R & EL & P/O \\
NIST SSDF& Data archiving & Q:Archive provenance and verification data? & M & PD & P/O \\
TUF & Record keeping & Sub-Q:Metadata with cryptographic hashes? & A & PD & P/O \\
TUF & Record keeping & Sub-Q:Record update start time? & R & PD & P/O \\
MDCF, CSA-STAR & Component registration & Sub-Q:Tool to track components and dependencies? & M & D & P/O \\
\hline
\end{tabular}
}
\vspace{0.5em}

\begin{tabular}{@{}ll@{}}
\textbf{Legend:} & \\
Lvl: & M = Mandatory, R = Recommended, A = Advanced \\
Phase: & PR = Preparation, Ac = Acquisition, D = Deployment, PD = Post-deployment, EL = Entire life cycle \\
Stakeholder: & P = Producer, C = Consumer, M = Manager, O = Operator \\
\bottomrule
\end{tabular}

\label{tab:checklist_example_traceability}
\end{table}

This category includes a total of 9 questions (5 questions and 4 sub-questions) which span all 3 implementation levels: mandatory, recommended, and advanced, demonstrating the layered nature of assurance practices required in this domain. Mandatory questions (e.g., SBOM registration, archival of verification data) establish essential controls for traceability and transparency, while recommended questions focus on strengthening ecosystem coordination and metadata management. Advanced controls, such as ensuring cryptographic integrity of metadata (e.g., from TUF), reflect a higher assurance posture suitable for systems with elevated risk exposure.

In terms of life cycle coverage, these items are distributed across preparation, deployment, and post-deployment phases, with some (e.g., documentation and record keeping) applicable to the entire life cycle. This illustrates the persistent and evolving nature of traceability concerns throughout software supply chains.

Stakeholder responsibilities are predominantly assigned to the producer group, reflecting their central role in generating, maintaining, and sharing traceability data. However, distinctions also exist; Managers are involved during preparation for governance and oversight, while operators are active in deployment and post-deployment for dissemination and validation of provenance information. This mapping supports role-specific accountability and targeted implementation across CI sector organisations.


\textbf{Key features.}
The checklist was designed and developed based on key features as follows.

\begin{itemize}
    \item Comprehensive: The checklist covers an entire software supply chain such as preparation, procurement/acquisition, development, deployment, and post-deployment. It supports various stakeholders throughout these phases and is designed to accommodate specific requirements related to CI sectors. Additionally, the checklist ensures that best practices are adhered to at each stage, facilitating a secure and efficient software delivery pipeline.
    \item Layered and connected: There are main questions (high-level questions) and sub-questions (low-level questions) addressing detailed implementation aspects. The varying levels of detail in the requirements and practices from the source frameworks have been structured into these layered questions. These questions are also grouped by security themes, which are strongly interconnected.
    \item Support maturity levels: The checklist provides different questions for companies at varying maturity levels: Mandatory, Recommended and Advanced. Mandatory questions are for companies at the foundational level, ensuring they meet essential security requirements. Recommended questions are for companies with a moderate level of maturity, guiding them toward best practices. Advanced questions are for companies at a high maturity level, challenging them to achieve and maintain industry-leading security standards.

    \item Beyond Checking, focusing on continuous improvement: Beyond simply identifying weaknesses or risk areas, the checklist goes a step further by connecting these findings with the reference architecture to recommend best practices for improvement. This approach ensures that security assessments are not just about compliance but are also an opportunity for continuous enhancement of processes and systems, driving the adoption of more robust and secure practices across the software supply chain.
\end{itemize}


\subsection{Related work}
Jaatun et al. \cite{Jaatun2024} conducted a survey-based compilation of good security practices to require of vendors in sectors like power distribution. Recognizing that critical systems rely heavily on third-party products, their checklist distills guidance from standards and industry best practices into concrete requirements for suppliers. An important insight from this work is the challenge of trust between infrastructure operators and vendors, where suppliers have misrepresented their security measures, underscoring the need for mechanisms to verify vendor claims. They suggest that critical infrastructure operators may need to employ more automated vendor auditing or collaborate (e.g. in procurement alliances) to enforce stringent security requirements on large suppliers. Tamanna et al. \cite{tamanna2024analyzing} investigated the adoption of the SLSA framework in real-world projects by analyzing over a thousand SLSA-related issues on GitHub. Their study identified significant barriers to effective implementation, revealing that many developers find SLSA challenging to implement correctly and struggle to understand its requirements across diverse ecosystems. The study highlights that widespread adoption of SLSA remains limited, primarily due to two key challenges: (1) the complexity of implementation and (2) unclear communication of the framework's guidelines. This work advances the understanding of framework usability, demonstrating that even well-intentioned frameworks like SLSA can falter without clear guidance and user-friendly practices—issues that are especially critical for resource-constrained teams, including those in sensitive sectors lacking specialized supply chain security expertise. Building on the theme of framework coverage, Hamer et al. \cite{hamer2025closingchain} examined multiple security frameworks and their effectiveness in mitigating real-world attack techniques. The researchers mapped the attack methods used in three notorious software supply chain incidents (e.g., SolarWinds, Log4j, and XZ Utils) to the defensive controls prescribed by ten different security frameworks (e.g., NIST SSDF, NIST 800-161, SLSA, and others). Their analysis revealed a sobering conclusion that critical security measures were missing from all of the frameworks examined. Specifically, they identified at least three essential mitigation tasks absent from every one of the ten frameworks, indicating that current guidance does not fully address certain attack vectors. The study concluded that even perfect compliance with all major frameworks would still leave organisations vulnerable to some supply chain attacks. 
Several studies\cite{melara2022what,Sammak2024Developers,EnckW22} have examined the practical challenges faced by developers and organizations in implementing secure supply chain practices. For example, Sammak et al.\cite{Sammak2024Developers} conducted an interview-based study, uncovering gaps between existing security guidelines and the specific needs of developers, highlighting the limitations of overly generalized software security regulations, guidelines and frameworks. 

\subsection{Threats to validity}

\textbf{Construct validity.}
The limited number of software supply chain security frameworks specifically tailored to critical infrastructure sectors poses the main threat to construct validity.
To mitigate this bias, we expanded our framework collection to include multiple complementary sources, such as widely adopted industry frameworks, sector-specific standards, and Australian software security frameworks and regulations, supplemented by a rapid review of relevant academic literature identified through Google Search. Together, these frameworks and academic studies form a broader and more representative knowledge base, supporting the development of robust, context-aware security controls that address both general and sector-specific risks in software supply chains.

\textbf{Internal validity.}
To support the integration of multiple frameworks, we systematically extracted security requirements and best practices from each source during the data extraction phase. These frameworks vary in their level of abstraction and granularity for security requirements. To ensure consistency during data analysis and synthesis, we defined a set of overarching categories and nested subcategories aligned with key phases in the software supply chain life cycle. The data extraction, synthesis, and classification processes were conducted by two researchers and validated through iterative discussions with both internal team members and external collaborators. Weekly meetings were held to ensure alignment and interpretive consistency.

\textbf{External validity.}
To strengthen external validity during the research process, we diversified our data sources to include internationally recognized software supply chain frameworks, Australian security frameworks and regulations, and relevant academic studies, thereby providing a balanced foundation that captures multiple governance and operational perspectives.
For each research question and the resulting checklist structure, we conducted multiple rounds of discussions with external industry and government partners to review, refine, and validate the interpretation of extracted requirements.
These collaborative consultations helped ensure that the checklist content aligns with real operational needs and sector-specific contexts.
Although large-scale pilot validation is still in progress, these collaborative processes have substantially enhanced the generalizability and practical relevance of our findings.
We consider continued stakeholder feedback and field adoption as essential next steps to further strengthen the external validity and long-term applicability of the proposed checklist.

\section{Conclusion}
This study presents a comprehensive analysis of software supply chain security frameworks and practices in the context of Australia's critical infrastructure sectors. By conducting a multivocal survey that integrates international, sector-specific frameworks, academic papers and Australian regulatory sources, we identified ten core categories of software security practices and mapped them across supply chain phases, stakeholder roles, and implementation levels. Building on these insights, we developed a structured, multi-layered checklist that enables organisations to assess, prioritise, and improve their software supply chain security posture. The checklist supports both baseline compliance and progressive maturity through its mandatory, recommended, and advanced question levels. Our findings reveal that existing frameworks remain fragmented and often lack contextual alignment with sector-specific operational needs, highlighting the need for more integrated and adaptable approaches. Our future work will focus on piloting the checklist with industry and government partners to evaluate its effectiveness, refine sector-specific profiles, and strengthen its practical adoption across diverse CI domains.







\bibliographystyle{unsrt}  
\bibliography{risk}

\end{document}